\documentclass[sigconf]{acmart}
\pdfoutput=1
\acmConference[IFL'19]{International Symposium on Implementation and Application of Functional Languages}{September 2019}{Singapore}
\acmYear{2020}
\copyrightyear{2020}
\acmDOI{} 
\startPage{1}

\setcopyright{none}

\usepackage{mathpartir} 
\usepackage{listings}

\lstdefinelanguage{IR} {
mathescape=true,
texcl=false,
morekeywords=[1] {
let, ret, proj, case, of, pap, ctor, inc, dec, reset, reuse, in
},
literate=
{→}{{\ensuremath{\rightarrow\,}}}1
{->}{{\ensuremath{\rightarrow\,}}}1
{_i}{{\ensuremath{_i}}}1
{_1}{{\ensuremath{_1}}}1
{_2}{{\ensuremath{_2}}}1
{_3}{{\ensuremath{_3}}}1
{_4}{{\ensuremath{_4}}}1
{_5}{{\ensuremath{_5}}}1
{₁}{{\ensuremath{_1}}}1
{₂}{{\ensuremath{_2}}}1
{₃}{{\ensuremath{_3}}}1
{₄}{{\ensuremath{_4}}}1
{₅}{{\ensuremath{_5}}}1
{:}{{\ensuremath{\,}}:{\ensuremath{\,}}}1
{;}{{\ensuremath{\,}};}1
{collecto}{{\ensuremath{\textit{collect}_\own{}}}}1
{₆}{{\ensuremath{_6}}}1,
identifierstyle={\rmfamily\itshape\color{black}},
keywordstyle=[1]{\ttfamily\bfseries\color{black}},
stringstyle=\ttfamily,
columns=fullflexible
}

\usepackage{amsthm} 
\usepackage{amssymb} 
\usepackage[capitalise]{cleveref} 
\usepackage{hyperref} 

\DeclareMathOperator{\dom}{dom}

\DeclareMathOperator{\dec}{dec}
\DeclareMathOperator{\inc}{inc}

\DeclareMathOperator{\FV}{FV}

\newcommand\RC{_{\textit{RC}}}
\newcommand\rcIR{\ensuremath{\lambda\RC}}
\def\Reset#1{\kw{reset}\ #1}
\def\Reuse_#1#2#3{\kw{reuse}\ #2\ \kw{in}\ \kw{ctor}_#1\ #3}
\def\Inc#1;#2{\kw{inc}\ #1;\ #2}
\def\Dec#1;#2{\kw{dec}\ #1;\ #2}

\newcommand{\var}[1]{\texttt{#1}}
\newcommand\abs[1]{\mid #1 \mid}

\newcommand\pure{_{\textit{pure}}}
\newcommand\pureIR{\ensuremath{\lambda\pure}}

\newcommand\ty[1]{\mathit{#1}}
\newcommand\syncat[1]{&\in \ty{#1}}
\newcommand\kw[1]{\texttt{\textbf{#1}}}
\newcommand\many[1]{\overline{#1}}
\newcommand\rcsem[5]{#1 \vdash \langle #2, #3 \rangle \Downarrow \langle #4, #5 \rangle}
\newcommand\nullptr{\rule{0.5em}{0.5em}}

\def\Pap#1#2{\kw{pap}\ #1\ #2}
\def\Ctor_#1#2{\kw{ctor}_#1\ #2}
\def\Proj_#1#2{\kw{proj}_#1\ #2}
\def\Ret#1{\kw{ret}\ #1}
\def\Let#1=#2;#3{\kw{let}\ #1 = #2;\ #3}
\def\Case#1#2{\kw{case}\ #1\ \kw{of}\ #2}
\def\Lambda#1.#2{\lambda\ #1.\ #2}

\newcommand\own{\ensuremath{\mathbb{O}}}
\newcommand\bor{\ensuremath{\mathbb{B}}}

\newcommand\reuse{_{\textit{reuse}}}

\title{Counting Immutable Beans}
\subtitle{Reference Counting Optimized for Purely Functional Programming}

\author{Sebastian Ullrich}
\affiliation{
  \institution{Karlsruhe Institute of Technology}
  \country{Germany}
}
\email{sebastian.ullrich@kit.edu}

\author{Leonardo de Moura}
\affiliation{
  \institution{Microsoft Research}           
  \country{USA}                   
}
\email{leonardo@microsoft.com}

\begin{abstract}
  Most functional languages rely on some kind of garbage collection
  for automatic memory management. They usually eschew reference
  counting in favor of a tracing garbage collector, which has less
  bookkeeping overhead at runtime. On the other hand, having
  an exact reference count of each value can enable
  optimizations such as destructive updates. We explore these
  optimization opportunities in the context of an eager, purely
  functional programming language. We propose a new mechanism for efficiently reclaiming
  memory used by nonshared values, reducing stress on the
  global memory allocator. We describe an approach for
  minimizing the number of reference counts updates using borrowed
  references and a heuristic for automatically inferring borrow
  annotations. We implemented all these techniques in a
  new compiler for an eager and purely functional programming language
  with support for multi-threading. Our preliminary experimental
  results demonstrate our approach is competitive and often
  outperforms state-of-the-art compilers.
\end{abstract}



\begin{document}
\maketitle

\section{Introduction}
\label{sec:intro}

Although reference counting~\cite{classicRC} (RC) is one of the oldest
memory management techniques in computer science, it is not considered
a serious garbage collection technique in the functional programming
community, and there is plenty of evidence it is in general inferior
to tracing garbage collection algorithms. Indeed, high-performance
compilers such as ocamlopt and GHC use tracing garbage
collectors. Nonetheless, implementations of several popular
programming languages, e.g., Swift, Objective-C, Python, and Perl, use
reference counting as a memory management technique. Reference
counting is often praised for its simplicity, but many disadvantages
are frequently reported in the
literature~\cite{GCbook,GCsurvey}. First, incrementing and
decrementing reference counts every time a reference is created or
destroyed can significantly impact performance because they not only
take time but also affect cache performance, especially in a
multi-threaded program~\cite{biasedRC}. Second, reference counting
cannot collect circular~\cite{cyclesRC} or self-referential
structures. Finally, in most reference counting implementations, pause
times are deterministic but may still be unbounded~\cite{BoehmCostRC}.

In this paper, we investigate whether reference counting is a
competitive memory management technique for purely functional
languages, and explore optimizations for reusing memory, performing
destructive updates, and for minimizing the number of reference count
increments and decrements. The former optimizations in particular are
beneficial for purely functional languages that otherwise can only
perform functional updates. When performing functional updates,
objects often die just before the creation of an object of the same
kind.  We observe a similar phenomenon when we insert a new element
into a pure functional data structure such as binary trees, when we
use \emph{map} to apply a given function to the elements of a list or
tree, when a compiler applies optimizations by transforming abstract
syntax trees, or when a proof assistant rewrites formulas. We call it the
\emph{resurrection hypothesis}: many objects die just before the
creation of an object of the same kind. Our new optimization takes
advantage of this hypothesis, and enables pure code to perform
destructive updates in all scenarios described above when objects are
not shared.  We implemented all the ideas reported here in the new
runtime and compiler for the Lean programming language~\cite{Lean}. We
also report preliminary experimental results that demonstrate our new
compiler produces competitive code that often outperforms the code
generated by high-performance compilers such as ocamlopt and GHC
(\cref{sec:evaluation}).

Lean implements a version of the Calculus of Inductive
Constructions~\cite{coquand:huet:88,coquand:paulin:mohring:90}, and it
has mainly been used as a proof assistant so far. Lean has a
metaprogramming framework for writing proof and code automation, where
users can extend Lean using Lean itself~\cite{ebner17meta}. Improving
the performance of Lean metaprograms was the primary motivation for the work
reported here, but one can apply the techniques reported here
to general-purpose functional programming languages.

We describe our approach as a series of refinements starting from
\pureIR{}, a simple intermediate representation for eager and purely
functional languages (\cref{sec:pure}). We remark that in Lean and
\pureIR{}, it is not possible to create cyclic data structures. Thus,
one of the main criticisms against reference counting does not
apply. From \pureIR{}, we obtain \rcIR{} by adding explicit instructions
for incrementing ($\kw{inc}$) and decrementing ($\kw{dec}$) reference
counts, and reusing memory (\cref{sec:rc}). The inspiration for explicit
RC instructions comes from the Swift compiler, as does the notion
of \emph{borrowed} references. In contrast to standard
(or \emph{owned}) references, of which there should be exactly as many
as the object's reference counter implies, borrowed references do not
update the reference counter but are \emph{assumed} to be kept alive
by a surrounding owned reference, further minimizing the number of
$\kw{inc}$ and $\kw{dec}$ instructions in generated code.

We present a simple compiler from \pureIR{} to \rcIR{},
discussing heuristics for inserting destructive updates, borrow
annotations, and \kw{inc}/\kw{dec} instructions (\cref{sec:compiler}).
Finally, we show that our approach is compatible with existing techniques
for performing destructive updates on array and string values, and
propose a simple and efficient approach for thread-safe reference
counting (\cref{sec:runtime}).

\emph{Contributions}. We present a reference counting scheme optimized for and
used by the next version of the Lean programming language.
\begin{itemize}
\item We describe how to reuse allocations in both user code and
  language primitives, and give a formal reference-counting semantics that
  can express this reuse.
\item We describe the optimization of using borrowed references.
\item We define a compiler that implements all these steps. The compiler is implemented
  in Lean itself and the source code is available.
\item We give a simple but effective scheme for avoiding atomic reference count
  updates in multi-threaded programs.
\item We compare the new Lean compiler incorporating these ideas with other
  compilers for functional languages and show its competitiveness.
\end{itemize}

\section{Examples}
\label{sec:examples}

In reference counting, each heap-allocated value contains a reference
count. We view this counter as a collection of tokens. The \kw{inc}
instruction creates a new token and \kw{dec} consumes it. When a function
takes an argument as an owned reference, it is responsible for
consuming one of its tokens. The function may consume the owned
reference not only by using the \kw{dec} instruction, but also by storing
it in a newly allocated heap value, returning it, or passing it to
another function that takes an owned reference. We illustrate our intermediate
representation (IR) and the use of owned and borrowed references with a series of
small examples.

The identity function $\textit{id}$ does not require any RC operation
when it takes its argument as an owned reference.
\begin{lstlisting}
id x = ret x
\end{lstlisting}
As another example, consider the function $\textit{mkPairOf}$ that takes $x$ and returns the pair $(x, x)$.
\begin{lstlisting}
mkPairOf x = inc x; let p = Pair x x; ret p
\end{lstlisting}
It requires an \kw{inc} instruction because two tokens for $x$ are consumed (we will also say that ``$x$ is consumed'' twice).
The function $\textit{fst}$ takes two arguments $x$ and $y$, and
returns $x$, and uses a $\kw{dec}$ instruction for consuming the unused $y$.
\begin{lstlisting}
fst x y = dec y; ret x
\end{lstlisting}
The examples above suggest that we do not need any RC operation when
we take arguments as owned references and consume them exactly
once. Now we contrast that with a function that only inspects its argument:
the function \textit{isNil} \textit{xs} returns
true if the list \textit{xs} is empty and false otherwise. If the argument $\textit{xs}$ is taken as an
owned reference, our compiler generates the following code
\begin{lstlisting}
  isNil xs = case xs of
    (Nil -> dec xs; ret true)
    (Cons -> dec xs; ret false)
\end{lstlisting}
We need the \kw{dec} instructions because a function must consume all
arguments taken as owned references.  One may notice that decrementing
$\text{xs}$ immediately after we inspect its constructor tag is
wasteful. Now assume that instead of taking the ownership of an RC
token, we could borrow it from the caller.  Then, the callee would not
need to consume the token using an explicit \kw{dec} operation. Moreover,
the caller would be responsible for keeping the borrowed value alive.
This is the essence of \emph{borrowed references}: a borrowed reference does not actually keep the referenced value
alive, but instead asserts that the value is kept alive by another, owned reference.
Thus, when \textit{xs} is a borrowed reference, we compile \textit{isNil} into our IR as
\begin{lstlisting}
  isNil xs = case xs of (Nil -> ret true) (Cons -> ret false)
\end{lstlisting}

As a less trivial example, we now consider the function \textit{hasNone} \textit{xs} that,
given a list of optional values, returns \emph{true} if \textit{xs} contains a \textit{None} value.
This function is often defined in a functional language as
\begin{lstlisting}
  hasNone []            = false
  hasNone (None : xs)   = true
  hasNone (Some x : xs) = hasNone xs
\end{lstlisting}
Similarly to \textit{isNil}, \textit{hasNone} only inspects its argument. Thus if \textit{xs} is taken
as a borrowed reference, our compiler produces the following RC-free IR code for it
\begin{lstlisting}
  hasNone xs = case xs of
    (Nil -> ret false)
    (Cons -> let h = proj$_{\textit{head}}$ xs; case h of
      (None -> ret true)
      (Some -> let t = proj$_{\textit{tail}}$ xs; let r = hasNone t; ret r))
\end{lstlisting}
Note that our \kw{case} operation does not introduce binders. Instead, we use explicit instructions $\kw{proj}_i$ for
accessing the head and tail of the \textit{Cons} cell. We use suggestive names
for cases and fields in these initial examples, but will later use indices instead.
Our borrowed inference heuristic discussed in \cref{sec:compiler} correctly tags $\textit{xs}$ as a borrowed parameter.

When using owned references, we know at run time whether a
value is shared or not simply by checking its reference counter.
We observed we could leverage this information and minimize
the amount of allocated and freed memory for constructor values such as a
list \textit{Cons} value. Thus, we have added two additional instructions to our IR:
$\kw{let}\ y = \kw{reset}\ x$ and $\kw{let}\ z = (\kw{reuse}\ y\ \kw{in}\ \kw{ctor}_i\ \many{w})$.
The two instructions are used together; if $x$ is a shared value, then $y$ is set to
a special reference $\nullptr$,
and the $\kw{reuse}$ instruction just allocates a new constructor value $\kw{ctor}_i\ \many{w}$.
If $x$ is not shared, then $\kw{reset}$ decrements the
reference counters of the components of $x$, and $y$ is set to $x$. Then, $\kw{reuse}$ reuses
the memory cell used by $x$ to store the constructor value $\kw{ctor}_i\ \many{w}$.
We illustrate these two instructions with the IR code for the list \textit{map} function
generated by our compiler as shown in \cref{sec:compiler}. The code uses our
actual, positional encoding of cases, constructors, and fields as described in
the next section.
\begin{lstlisting}
  map f xs = case xs of
    (ret xs)
    (let x = proj_1 xs; inc x; let s = proj_2 xs; inc s;
     let w = reset xs;
     let y = f x; let ys = map f s;
     let r = (reuse w in ctor_2 y ys); ret r)
\end{lstlisting}
We remark that if the list referenced by $\textit{xs}$ is not shared, the code above
does not allocate any memory. Moreover, if \textit{xs} is a nonshared list of list of integers,
then $\textit{map}\ (\textit{map}\ \textit{inc})\ \textit{xs}$ will not allocate
any memory either.
This example also demonstrates it is not a good idea, in general, to fuse
$\kw{reset}$ and $\kw{reuse}$ into a single instruction: if we removed the
$\kw{let}\ w = \kw{reset}\ \textit{xs}$ instruction and directly used
\textit{xs} in \kw{reuse},
then when we execute the recursive application $\textit{map}\ f\ s$, the reference counter for $s$ would
be greater than $1$ even if the reference counter for $\textit{xs}$ was 1.
We would have a reference from $\textit{xs}$ and another from $s$, and memory reuse would not occur in the recursive applications.
Note that removing the $\kw{inc}\ s$ instruction is incorrect when
$\textit{xs}$ is a shared value.
Although the $\kw{reset}$ and $\kw{reuse}$ instructions can in general be used for reusing memory
between two otherwise unrelated values, in examples like \textit{map} where the
reused value has a close semantic connection to the reusing value, we will use
common functional vocabulary and say that the list is being \emph{destructively updated} (up to the first shared cell).

As another example, a zipper is a technique for traversing and efficiently updating data structures, and it is particularly useful for
purely functional languages. For example, the list zipper is a pair of lists, and it allows one to
move forward and backward, and to update the current position.
The \textit{goForward} function is often defined as
\begin{lstlisting}
   goForward ([], bs)   = ([], bs)
   goForward (x : xs, bs) = (xs, x : bs)
\end{lstlisting}
In most functional programming languages, the second equation allocates a
new pair and \textit{Cons} value.  
The functions \textit{map} and \textit{goForward} both satisfy our
resurrection hypothesis. Moreover, the result of a \textit{goForward}
application is often fed into another \textit{goForward} or
\textit{goBackward} application.  Even if the initial value was shared,
every subsequent application takes a nonshared pair, and
memory allocations are avoided by the code produced by our compiler.
\begin{lstlisting}
   goForward p = case p of
     (let xs = proj_1 p; inc xs;
      case xs of
        (ret p)
        (let bs = proj_2 p; inc bs;
         let c_1 = reset p;
         let x = proj_1 xs; inc x; xs' = proj_2 xs; inc xs';
         let c_2 = reset xs;
         let bs' = (reuse c_2 in ctor_2 x bs);
         let r   = (reuse c_1 in ctor_1 xs' bs'); ret r))
\end{lstlisting}

\section{The pure IR}
\label{sec:pure}

Our source language \pureIR{} is a simple untyped functional
intermediate representation (IR) in the style of A-normal
form~\cite{Flanagan:1993:ECC:155090.155113}. It captures the relevant
features of the actual IR we have implemented and avoids unnecessary
complexity that would only distract the reader from the ideas proposed here.
\begin{alignat*}{3}
  w, x, y, z \syncat{Var}\\
  c \syncat{Const}\\
  e \syncat{Expr} &::=&\ c\ \many{y} \mid \Pap c \many{y} \mid x\ y \mid \Ctor_i \many{y} \mid \Proj_i x \\
  F \syncat{FnBody} &::=&\ \Ret x \mid \Let x = e; F \mid \Case x \many{F} \\
  f \syncat{Fn} &::=&\ \Lambda \many{y}. F \\
  \delta \syncat{Program} &=&\ \ty{Const} \rightharpoonup \ty{Fn}
\end{alignat*}
All arguments of function applications are variables. The applied function is
a constant $c$, with partial applications marked with the keyword \kw{pap}, a
variable $x$, the $i$-th constructor of an
erased datatype, or the special function $\kw{proj}_i$, which returns the $i$-th
argument of a constructor application. Function bodies always end with
evaluating and returning a variable.
They can be chained with (non-recursive) \kw{let} statements and branch using \kw{case}
statements, which evaluate to their $i$-th arm given an application of $\kw{ctor}_i$.
As further detailed in \cref{sec:incdec}, we consider tail calls to be of the form
$\kw{let}\ r = c\ \many{x};\ \kw{ret}\ r$.
A program is a partial map from constant names to their implementations. The
body of a constant's implementation may refer back to the constant,
which we use to represent recursion, and analogously mutual recursion.
In examples, we use $f\ \many{x} = F$ as syntax sugar for $\delta(f) = \lambda\ \many{x}.\ F$.

As an intermediate representation, we can and should impose restrictions on the
structure of \pureIR{} to simplify working with it. We assume that

\begin{itemize}
\item all constructor applications are fully applied by eta-expanding them.
\item no constant applications are over-applied by splitting them into two applications where necessary.
\item all variable applications take only one argument, again by splitting them
  where necessary. While this simplification can introduce additional
  allocations of intermediary partial applications, it greatly simplifies the
  presentation of our operational semantics. All presented program
  transformations can be readily extended to a system with $n$-ary variable
  applications, which are handled analogously to $n$-ary constant applications.
\item every function abstraction has been lambda-lifted
  to a top-level constant $c$.
\item trivial bindings $\kw{let}\ x = y$ have been eliminated through copy propagation.
\item all dead \kw{let} bindings have been removed.
  \item all parameter and \kw{let} names of a function are mutually distinct. Thus we
    do not have to worry about name capture.
\end{itemize}

In the actual IR we have implemented\footnote{\url{https://github.com/leanprover/lean4/blob/IFL19/library/init/lean/compiler/ir/basic.lean}}, we also have instructions for
storing and accessing unboxed data in constructor values, boxing
and unboxing machine integers and scalar values, and creating literals
of primitive types such as strings and numbers. Our IR also supports
\emph{join points} similar to the ones used in the Haskell Core
language~\cite{JoinPoints}. Join points are local
function declarations that are never partially applied (i.e., they never
occur in $\kw{pap}$ instructions), and are always tail-called. The actual
IR has support for defining join points, and a $\kw{jmp}$ instruction for
invoking them.

\section{Syntax and semantics of the reference-counted IR}
\label{sec:rc}

The target language \rcIR{} is an extension of \pureIR{}:
\begin{alignat*}{3}
  e \syncat{Expr} &&::= \ldots \mid \Reset x \mid \Reuse_i x \many{y} \\
  F \syncat{FnBody} &&::= \ldots \mid \Inc x; F \mid \Dec x; F
\end{alignat*}
We use the subscripts $\pure$ or $\RC$ (e.g., $\ty{Expr}\pure$ or $\ty{Expr}\RC$) to refer to the base or
extended syntax, respectively, where otherwise ambiguous.
The new expressions \kw{reset} and \kw{reuse} work together to reuse memory used
to store constructor values, and, as discussed in \cref{sec:examples}, simulate
destructive updates in constructor values.


We define the semantics of \rcIR{} (Figures~\ref{fig:rc:sem1}~and~\ref{fig:rc:sem2}) using a
big-step relation $\rcsem{\rho}{F}{\sigma}{l}{\sigma'}$ that maps the
body $F$ and a mutable \emph{heap} $\sigma$ under a context $\rho$ to
a location and the resulting heap. The context $\rho$ maps
variables to locations. A heap $\sigma$ is a mapping from locations to
pairs of values and reference counters.
A value is a constructor value
or a partially-applied constant.
The reference counters of live values should always be positive; dead values
are removed from the heap map.
\begin{alignat*}{3}
  l &\in \ty{Loc}\\
  \rho &\in \ty{Ctxt} &=&\ \ty{Var} \rightharpoonup \ty{Loc}\\
  \sigma &\in \ty{Heap} &=&\ \ty{Loc} \rightharpoonup \ty{Value} \times \mathbb{N}^+\\
  v \syncat{Value} ::&=&\ \kw{ctor}_i\ \many{l} \mid \kw{pap}\ c\ \many{l}
\end{alignat*}
When applying a variable, we have to be careful to increment the
partial application arguments when copying them out of the \kw{pap} cell, and to
decrement the cell afterwards.\footnote{If the \kw{pap} reference is unique, the
two steps can be coalesced so that the arguments do not have to be touched.}
We cannot do so via explicit reference counting instructions because the number of
arguments in a \kw{pap} cell is not known statically.
\begin{figure}
\begin{gather*}
  \infer[Const-App-Full]{
    \delta(c) = \lambda\ \many{y_{\textit{c}}}.\ F \\
    \many{l} = \many{\rho(y)} \\
    \rcsem{[\many{y_{\textit{c}}} \mapsto \many{l}]}{F}{\sigma}{l'}{\sigma'}
  }{\rcsem{\rho}{c\ \many{y}}{\sigma}{l'}{\sigma'}} \\
  \infer[Const-App-Part]{
    \delta(c) = \lambda\ \many{y_{\textit{c}}}.\ F \\
    \many{l} = \many{\rho(y)} \\
    \abs{\many{l}} < \abs{\many{y_c}} \\
    l' \not\in \dom(\sigma)
  }{\rcsem{\rho}{\kw{pap}\ c\ \many{y}}{\sigma}{l'}{\sigma[l' \mapsto (\kw{pap}\ c\ \many{l}, 1)]}} \\
  \infer[Var-App-Full]{
    \sigma(\rho(x)) = (\kw{pap}\ c\ \many{l}, \_) \\
    \delta(c) = \lambda\ \many{y_c}.\ F \\
    l_y = \rho(y) \\
    \rcsem{[\many{y_c} \mapsto \many{l}\ l_y]}{F}{\dec(\rho(x), \inc(\many{l}, \sigma))}{l'}{\sigma'}
  }{\rcsem{\rho}{x\ y}{\sigma}{l'}{\sigma'}} \\
  \infer[Var-App-Part]{
    \sigma(\rho(x)) = (\kw{pap}\ c\ \many{l}, \_) \\
    \delta(c) = \lambda\ \many{y_c}.\ F \\
    l_y = \rho(y) \\
    \abs{\many{l}\ l_y} < \abs{\many{y_c}} \\
    l' \not\in \dom(\sigma)
  }{\rcsem{\rho}{x\ y}{\sigma}{l'}{\dec(\rho(x), \inc(\many{l}, \sigma))[l' \mapsto (\kw{pap}\ c\ \many{l}\ l_y, 1)]}} \\
  \infer[Ctor-App]{
    \many{l} = \many{\rho(y)} \\
    l' \not\in \dom(\sigma)
  }{\rcsem{\rho}{\kw{ctor}_i\ \many{y}}{\sigma}{l'}{\sigma[l' \mapsto (\kw{ctor}_i\ \many{l}, 1)]}}\;\; \\
  \infer[Proj]{
    \sigma(\rho(x)) = (\kw{ctor}_j\ \many{l}, \_) \\
    l' = \many{l}_i
  }{\rcsem{\rho}{\kw{proj}_i\ x}{\sigma}{l'}{\sigma}}\;\;
  \infer[Return]{
    \rho(x) = l
  }{\rcsem{\rho}{\kw{ret}\ x}{\sigma}{l}{\sigma}} \\
  \infer[Let]{
    \rcsem{\rho}{e}{\sigma}{l}{\sigma'} \\
    \rcsem{\rho[x \mapsto l]}{F}{\sigma'}{l'}{\sigma''}
  }{\rcsem{\rho}{\kw{let}\ x = e;\ F}{\sigma}{l'}{\sigma''}} \\
  \infer[Case]{
    \sigma(\rho(x)) = (\kw{ctor}_i\ \many{l}, \_) \\
    \rcsem{\rho}{F_i}{\sigma}{l'}{\sigma'}
  }{\rcsem{\rho}{\kw{case}\ x\ \kw{of}\ \many{F}}{\sigma}{l'}{\sigma'}}
\end{gather*}
\caption{\rcIR{} semantics\label{fig:rc:sem1}: the \pureIR{} fragment}
\end{figure}

\begin{figure}
\begin{gather*}
  \infer[Inc]{\rcsem{\rho}{F}{\inc(\rho(x), \sigma)}{l'}{\sigma'}}{\rcsem{\rho}{\kw{inc}\ x;\ F}{\sigma}{l'}{\sigma'}}\;\;
  \infer[Dec]{
    \rcsem{\rho}{F}{\dec(\rho(x), \sigma)}{l'}{\sigma'}
  }{\rcsem{\rho}{\kw{dec}\ x;\ F}{\sigma}{l'}{\sigma'}}
\end{gather*}
\begin{align*}
  \inc(l, \sigma) &= \sigma[l \mapsto (v, i + 1)] \;\;\;\text{ if } \sigma(l) = (v, i) \\
  \inc(l\ \many{l'}, \sigma) &= \inc(\many{l'}, \inc(l, \sigma)) \\
  \dec(l, \sigma) &= \begin{cases}
    \sigma &\text{if } l = \nullptr \\
    \sigma[l \mapsto (v, i - 1)] &\text{if } \sigma(l) = (v, i), i > 1\\
    \dec(\many{l'}, \sigma[l \mapsto \bot]) &\text{if } \sigma(l) = (\kw{pap}\ c\ \many{l'}, 1)\\
    \dec(\many{l'}, \sigma[l \mapsto \bot]) &\text{if } \sigma(l) = (\kw{ctor}_i\ \many{l'}, 1)
  \end{cases}\\
  \dec(l\ \many{l'}, \sigma) &= \dec(\many{l'}, \dec(l, \sigma))
\end{align*}
\begin{gather*}
  \infer[Reset-Uniq]{
    \rho(x) = l \\
    \sigma(l) = (\kw{ctor}_i\ \many{l'}, 1)
  }{\rcsem{\rho}{\kw{reset}\ x}{\sigma}{l}{\dec(\many{l'}, \sigma[l \mapsto (\kw{ctor}_i\ \nullptr^{\abs{\many{l'}}}, 1)])}} \\
  \infer[Reset-Shared]{
    \rho(x) = l \\
    \sigma(l) = (\_, i) \\
    i \neq 1
  }{\rcsem{\rho}{\kw{reset}\ x}{\sigma}{\nullptr}{\dec(l, \sigma)}} \\
  \infer[Reuse-Uniq]{
    \rho(x) = l \\
    \sigma(l) = (\kw{ctor}_j\ \nullptr^{\abs{\many{y}}}, 1) \\
    \many{\rho(y)} = \many{l''} \\
  }{\rcsem{\rho}{\kw{reuse}\ x\ \kw{in}\ \kw{ctor}_i\ \many{y}}{\sigma}{l}{\sigma[l \mapsto (\kw{ctor}_i\ \many{l''}, 1)]}} \\
  \infer[Reuse-Shared]{
    \rho(x) = \nullptr \\
    \rcsem{\rho}{\kw{ctor}_i\ \many{y}}{\sigma}{l'}{\sigma'}
  }{\rcsem{\rho}{\kw{reuse}\ x\ \kw{in}\ \kw{ctor}_i\ \many{y}}{\sigma}{l'}{\sigma'}}
\end{gather*}
\vspace{-1.5em}
\caption{\rcIR{} semantics cont.}
\label{fig:rc:sem2}
\end{figure}
Decrementing a unique reference removes the value from the heap and recursively
decrements its components.
\kw{reset}, when used on a unique reference, eagerly decrements the components
of the referenced value,
replaces them with $\nullptr$,\footnote{which can be represented by
  any unused pointer value such as the null pointer in a real implementation.
  In our actual implementation, we avoid this memory write by introducing a
  \texttt{del} instruction that behaves like \texttt{dec} but ignores the
  constructor fields.}
and returns the location of the now-invalid cell. This value is intended to be
used only by \kw{reuse} or \kw{dec}. The former reuses it for a new constructor
cell, asserting that its size is compatible with the old cell. The latter frees
the cell, ignoring the replaced children.

If \kw{reset} is used on a shared, non-reusable reference, it behaves like
\kw{dec} and returns $\nullptr$, which instructs \kw{reuse} to behave like
\kw{ctor}. Note that we cannot simply return the reference in both cases and do
another uniqueness check in \kw{reuse} because other code between the two
expressions may have altered its reference count.



\section{A compiler from \pureIR{} to \rcIR{}}
\label{sec:compiler}

Following the actual implementation of our compiler, we will discuss a 
compiler from \pureIR{} to \rcIR{} in three steps:

\begin{enumerate}
\item Inserting \kw{reset}/\kw{reuse} pairs (\cref{sec:update})
\item Inferring borrowed parameters (\cref{sec:borrowinf})
\item Inserting \kw{inc}/\kw{dec} instructions (\cref{sec:incdec})
\end{enumerate}

The first two steps are optional for obtaining correct \rcIR{} programs.

\subsection{Inserting destructive update operations}
\label{sec:update}

In this subsection, we will discuss a heuristics-based implementation of a function
\[ \delta\reuse : \ty{Const} \rightarrow \ty{Fn}_{\textit{RC}} \]
that inserts \kw{reset}/\kw{reuse} instructions.
Given $\kw{let}\ z = \kw{reset}\ x$, we remark that, in every control path, $z$ may
appear at most once, and in one of the following two instructions:
$\kw{let}\ y = \kw{reuse}\ z\ \kw{ctor}_i\ \many{w}$, or $\kw{dec}\ z$.
We use $\kw{dec}\ z$ for control paths where $z$ cannot be reused.
We implement the function $\delta\reuse$ as
\[ \delta\reuse(c) = \lambda\ \many{y}.\ R(F) \text{ where } \delta(c) = \lambda\ \many{y}.\ F \]
The function $R(F)$ (\cref{fig:resetreuse}) uses a simple heuristic for replacing
$\kw{ctor}_i\ \many{y}$ expressions occurring in $F$ with
$\kw{reuse}\ w\ \kw{in ctor}_i\ \many{y}$ where $w$ is a fresh
variable introduced by $R$ as the result of a new $\kw{reset}$
operation. For each arm $F_i$ in
a $\kw{case}\ x\ \kw{of}\ \many{F}$ operation, the function $R$ requires the arity $n$
of the corresponding matched constructor. In the actual implementation,
we store this information for each arm when we compile our typed frontend language into \pureIR{}.
The auxiliary functions $D$ and $S$ implement the \emph{dead} variable search and \emph{substitution} steps respectively.
For each \kw{case} operation, $R$ attempts to insert \kw{reset}/\kw{reuse} instructions
for the variable matched by the \kw{case}. This is done using $D$ in each arm of the \kw{case}.
Function $D(z, n, F)$ takes as parameters the variable $z$ to reuse and the arity $n$ of the matched constructor.
$D$ proceeds to the first location where $z$ is dead, i.e. not used in the remaining function body,
and then uses $S$ to attempt to find and substitute a \emph{matching}
constructor $\kw{ctor}_i\ \many{y}$ instruction with a $\kw{reuse}\ w\ \kw{in}\ \kw{ctor}_i\ \many{y}$ in the remaining code.
If no matching constructor instruction can be found, $D$ does not modify the function body.
\begin{figure}
  \begin{flalign*}
    &R : \ty{FnBody}\pure \rightarrow \ty{FnBody}\RC \\
    &R(\kw{let}\ x = e;\ F) = \kw{let}\ x = e;\ R(F) \\
    &R(\kw{ret}\ x) = \kw{ret}\ x \\
    &R(\kw{case}\ x\ \kw{of}\ \many{F}) = \kw{case}\ x\ \kw{of}\ \many{D(x, n_i, R(F_i))} && \\
      &\phantom{==} \text{where } n_i = \#\text{fields of } x \text{ in } i\text{-th branch}
  \end{flalign*}

\begin{flalign*}
  &D : \ty{Var} \times \mathbb{N} \times \ty{FnBody}\RC \rightarrow \ty{FnBody}\RC && \\
  &D(z, n, \kw{case}\ x\ \kw{of}\ \many{F}) = \kw{case}\ x\ \kw{of}\ \many{D(z, n, F)} \\
  &D(z, n, \kw{ret}\ x) = \kw{ret}\ x \\
  &D(z, n, \kw{let}\ x = e;\ F) = \kw{let}\ x = e;\ D(z, n, F) \\
  &\phantom{==} \text{if } z \in e \text{ or } z \in F \\
  &D(z, n, F) = \kw{let}\ w = \kw{reset}\ z;\ S(w, n, F) \\
  &\phantom{==} \text{otherwise, if } S(w, n, F) \ne F \text{ for a fresh }w \\
  &D(z, n, F) = F \phantom{=} \text{otherwise}
\end{flalign*}

\begin{flalign*}
  &S : \ty{Var} \times \mathbb{N} \times \ty{FnBody}\RC \rightarrow \ty{FnBody}\RC && \\
  &S(w, n, \kw{let}\ x = \kw{ctor}_i\ \many{y};\ F) = \kw{let}\ x = \kw{reuse}\ w\ \kw{in}\ \kw{ctor}_i\ \many{y};\ F \\
  &   \phantom{==} \text{if } \abs{\many{y}} = n \\
  &S(w, n, \kw{let}\ x = e;\ F) = \kw{let}\ x = e;\ S(w, n, F) \phantom{=} \text{otherwise} \\
  &S(w, n, \kw{ret}\ x) = \kw{ret}\ x \\
  &S(w, n, \kw{case}\ x\ \kw{of}\ \many{F}) = \kw{case}\ x\ \kw{of}\ \many{S(w, n, F)}
\end{flalign*}
\caption{Inserting \kw{reset}/\kw{reuse} pairs}
\label{fig:resetreuse}
\end{figure}

As an example, consider the $\textit{map}$ function for lists
\begin{lstlisting}
  map f xs = case xs of
    (ret xs)
    (let x = proj_1 xs; let s := proj_2 xs;
     let y = f x; let ys = map f s;
     let r = ctor_2 y ys; ret r)
\end{lstlisting}
Applying $R$ to the body of $\textit{map}$ , we have $D$ looking for opportunities to \kw{reset}/\kw{reuse} $\textit{xs}$
in both \kw{case} arms.
Since $\textit{xs}$ is unused after $\kw{let}\ s = \kw{proj}_2\ \textit{xs}$, $S$ is applied to the rest of the function,
looking for constructor calls with two parameters. Indeed, such a call can be found in the let-binding for $r$. Thus,
function $D$ successfully inserts the appropriate instructions, and we obtain the function described in \cref{sec:examples}.
Now, consider the list \emph{swap} function that swaps the first two elements of a list.
It is often defined as
\begin{lstlisting}
  swap [] = []
  swap [x] = [x]
  swap (x:y:zs) = y:x:zs
\end{lstlisting}
In \pureIR{}, this function is encoded as
\begin{lstlisting}
  swap xs = case xs of
     (ret xs)
     (let t_1 = proj_2 xs; case t_1 of
       (ret xs)
       (let h_1 = proj_1 xs;
        let h_2 = proj_1 t_1; let t_2 = proj_2 t_1;
        let r_1 = ctor_2 h_1 t_2; let r_2 = ctor_2 h_2 r_1; ret r_2))
\end{lstlisting}
By applying $R$ to \emph{swap}, we obtain
\pagebreak
\begin{lstlisting}
  swap xs = case xs of
    (ret xs)
    (let t_1 = proj_2 xs; case t_1 of
      (ret xs)
      (let h_1 = proj_1 xs; let w_1 = reset xs;
       let h_2 = proj_1 t_1; let t_2 = proj_2 t_1;
       let w_2 = reset t_1; let r_1 = reuse w_2 in ctor_2 h_1 t_2;
       let r_2 = reuse w_1 in ctor_2 h_2 r_1; ret r_2))
\end{lstlisting}
Similarly to the $\textit{map}$ function, the code generated for the
function $\textit{swap}$ will \emph{not} allocate any memory when the
list value is not shared. This example demonstrates that our heuristic
procedure can avoid memory allocations even in functions containing
many nested $\kw{case}$ instructions. The example also makes it clear
that we could further optimize our \rcIR{} by adding additional
instructions. For example, we can
add an instruction that combines $\kw{reset}$ and $\kw{reuse}$ into a single instruction
and is used in situations where $\kw{reuse}$ occurs \emph{immediately}
after the corresponding $\kw{reset}$ instruction such as in the example above where we have
$\kw{let}\ w_2 = \kw{reset}\ t_1;\ \kw{let}\ r_1 =\kw{reuse}\ w_2\ \kw{in}\ \kw{ctor}_2\ h_1\ t_2$,





\subsection{Inferring borrowing signatures}
\label{sec:borrowinf}

\label{sec:beta}
We now consider the problem of inferring borrowing signatures, i.e. a mapping
$\beta : \ty{Const} \rightharpoonup \{\own{}, \bor{}\}^*$, which for every
function should return a list describing each parameter of the function as
either \own{}{}wned or \bor{}{}orrowed.
Borrow annotations can be provided manually by users (which is always safe), but we have two motivations
for inferring them: avoiding the burden of annotations, and making our IR a convenient target
for other systems (e.g., Coq, Idris, and Agda) that do not have borrow annotations.

If a function $f$ takes a parameter $x$ as a borrowed reference, then
at runtime $x$ may be a shared value even when its reference counter
is $1$. Thus, we must never mark $x$ as borrowed if it is used by
a $\kw{let}\ y = \kw{reset}\ x$ instruction. We also assume that
each $\beta(c)$ has the same length as the corresponding parameter
list in $\delta(c)$.

Partially applying constants with borrowed parameters is also
problematic because, in general, we cannot statically assert that the
resulting value will not escape the current function and thus the
scope of borrowed references. Therefore we extend $\delta\reuse$ to
the program $\delta_\beta$ by defining a trivial wrapper
constant $c_\own{} := c$ (we will assume that this name is fresh) for any
such constant $c$, set $\beta(c_\own{}) := \many{\own{}}$, and replace any
occurrence of $\Pap c \many{y}$ with $\Pap {c_\own{}} \many{y}$.
The compiler step given in the next subsection will,
as part of the general transformation,
insert the necessary \kw{inc} and \kw{dec} instructions into
$c_\own{}$ to convert between the two signatures.

Our heuristic is based on the fact that when we mark a parameter as
borrowed, we reduce the number of RC operations needed, but we
also prevent $\kw{reset}$ and $\kw{reuse}$ as well as primitive
operations from reusing memory cells.  Our heuristic collects which
parameters and variables should be owned.  We say a parameter $x$
should be owned if $x$ or one of its projections is used in a
\kw{reset}, or is passed to a function that takes an owned reference.
The latter condition is a heuristic and is not required for
correctness.  We use it because the function taking an owned reference
may try to reuse its memory cell.  A formal definition is given in
\cref{fig:collect_O}. Many refinements are possible, and we
discuss one of them in the next section.
\begin{figure}
\begin{lstlisting}
$\textit{collect}_\own{}$ : $\ty{FnBody}\RC \rightarrow 2^{\ty{Vars}}$
$\textit{collect}_\own{}$(let z = ctor_i $\many{x}$; F) = $\textit{collect}_\own{}(F)$
$\textit{collect}_\own{}$(let z = reset x; F) = $\textit{collect}_\own{}(F) \cup \{x\}$
$\textit{collect}_\own{}$(let z = reuse x in ctor_i $\many{x}$; F) = $\textit{collect}_\own{}(F)$
$\textit{collect}_\own{}$(let z = c $\many{x}$; F) = $\textit{collect}_\own{}(F) \cup \{x_i \in \many{x}\ |\ \beta(c)_i = \own{} \}$
$\textit{collect}_\own{}$(let z = x y; F) = $\textit{collect}_\own{}(F) \cup \{x, y\}$
$\textit{collect}_\own{}$(let z = pap $c_\own{}$ $\many{x}; F$) = $\textit{collect}_\own{}(F) \cup \{\many{x}\}$
$\textit{collect}_\own{}$(let z = proj_i x; F) = $\textit{collect}_\own{}(F) \cup \{x\}$ $\text{if } z \in \textit{collect}_\own{}(F)$
$\textit{collect}_\own{}$(let z = proj_i x; F) = $\textit{collect}_\own{}(F)$ $\text{if } z \not\in \textit{collect}_\own{}(F)$
$\textit{collect}_\own{}$(ret x) = $\emptyset$
$\textit{collect}_\own{}$(case x of $\many{F}$) = $\bigcup_{F_i \in \many{F}} \textit{collect}_\own{}(F_i)$
\end{lstlisting}
\caption{Collecting variables that should not be marked as borrowed\label{fig:collect_O}}
\end{figure}
Note that if a call is recursive, we do not know which parameters are owned, yet. Thus, given
$\delta(c) = \lambda \many{y}.\ b$, we infer the value of $\beta(c)$ by starting
with the approximation $\beta(c) = \bor{}^n$, then we compute $S = \textit{collect}_\own{}(b)$,
update $\beta(c)_i := \own{}$ if $y_i \in S$, and
repeat the process until we reach a fix point and no further updates
are performed on $\beta(c)$. The procedure described here does not
consider mutually recursive definitions, but this is a simple
extension where we process a block of mutually recursive functions
simultaneously.
By applying our heuristic to the $\textit{hasNone}$ function described before, we obtain $\beta(\textit{hasNone}) = \bor{}$. That is, in an application $\textit{hasNone}\ \textit{xs}$, $\textit{xs}$ is taken as a borrowed reference.


\subsection{Inserting reference counting operations}
\label{sec:incdec}

Given any well-formed definition of $\beta$ and $\delta_\beta$, we
finally give a procedure for correctly inserting \kw{inc} and \kw{dec}
instructions.\footnote{We will tersely say that a variable $x$ ``is incremented/decremented'' when an \kw{inc}/\kw{dec}
  operation is applied to it, i.e.\ the RC of the referenced object is incremented/decremented at runtime.}
\begin{align*}
  \delta_{\textit{RC}}(c) &: \ty{Const} \rightarrow \ty{Fn}_{\textit{RC}} \\
  \delta_{\textit{RC}}(c) &= \lambda\ \many{y}.\ \own{}^-(\many{y}, C(F, \beta_l)) &&\text{ where } \hspace{-1em}&&\delta_\beta(c) = \lambda\ \many{y}.\ F, \\
  &&&&&\beta_l = [\many{y} \mapsto \beta(c), \ldots \mapsto \own{}]
\end{align*}
The map $\beta_l : \ty{Var} \rightarrow \{\own{}, \bor{}\}$
keeps track of the borrow status of each local variable. For simplicity, we default all missing entries to \own{}.

In general, variables should be incremented prior to being used in an \emph{owned context}
that consumes an RC token.
Variables used in any other (\emph{borrowed}) context do not need to be
incremented. Owned references should be decremented after their last use. We use
the following two helper functions to conditionally add RC instructions
(\cref{fig:compiler}) in these contexts:
\begin{itemize}
\item $\own{}^+_x$ prepares $x$ for usage in an owned context by incrementing it. The
  increment can be omitted on the last use of an owned variable, with $V$
  representing the set of live variables after the use.
  \begin{align*}
    \own{}^+_x(V, F, \beta_l) &= F &&\text{if } \beta_l(x) = \own{} \wedge x \not\in V \\
    \own{}^+_x(V, F, \beta_l) &= \kw{inc}\ x;\ F &&\text{otherwise}
  \end{align*}
\item $\own{}^-_x$ decrements $x$ if it is both owned and dead.
  $\own{}^-(\many{x}, F, \beta_l)$ decrements multiple variables, which may be needed at
  the start of a function or \kw{case} branch.
  \begin{lstlisting}
    $\own{}^-_x(F, \beta_l) = \kw{dec}\ x;\ F$  $\hspace{0.8cm} \text{if } \beta_l(x) = \own{} \wedge x \not\in \FV(F)$
    $\own{}^-_x(F, \beta_l) = F$  $\hspace{1.7cm} \text{otherwise}$
    $\own{}^-(x\ \many{x'}, F, \beta_l) = \own{}^-(\many{x'}, \own{}^-_x(F, \beta_l), \beta_l)$
    $\own{}^-([], F, \beta_l) = F$
  \end{lstlisting}
\end{itemize}
Applications are handled separately, recursing over the arguments and parameter
borrow annotations in parallel (in the case of partial applications, the former
list may be shorter); for partial, variable and constructor applications, the
latter default to \own{}.
\begin{figure}
\begin{lstlisting}
C : $\ty{FnBody}_{\textit{RC}} \times (\ty{Var} \rightarrow \{\own{}, \bor{}\}) \rightarrow \ty{FnBody}_{\textit{RC}}$
C (ret x, $\beta_l$) = $\own{}^+_x$($\emptyset$, ret x, $\beta_l$)
C (case x of $\many{F}$, $\beta_l$) = case x of $\many{\own{}^-(\many{y}, C(F, \beta_l), \beta_l)}$
                   $\text{where } \{\many{y}\} = \FV(\Case x \many{F})$
C (let y = proj_i x; F, $\beta_l$) = let y = proj_i x; inc y; $\own{}^-_x(C(F, \beta_l), \beta_l)$
                         $\text{if } \beta_l(x) = \own{}$
C (let y = proj_i x; F, $\beta_l$) = let y = proj_i x; C (F, $\beta_l[y \mapsto \bor{}]$)
                         $\text{if } \beta_l(x) = \bor{}$
C (let y = reset x; F, $\beta_l$) = let y = reset x; C (F, $\beta_l$)
C (let z = c $\many{y}$; F, $\beta_l$) = $C_{\textit{app}}(\many{y}, \beta(c), \kw{let}\ z = c\ \many{y};\ C(F, \beta_l), \beta_l)$
C (let z = pap c $\many{y}$; F, $\beta_l$) = $C_{\textit{app}}(\many{y}, \beta(c), \kw{let}\ z = \kw{pap}\ c\ \many{y};\ C(F, \beta_l), \beta_l)$
C (let z = x y; F, $\beta_l$) = $C_{\textit{app}}(x\ y, \own{}\ \own{}, \kw{let}\ z = x\ y;\ C(F, \beta_l), \beta_l)$
C (let z = ctor_i $\many{y}$; F, $\beta_l$) = $C_{\textit{app}}(\many{y}, \many{\own{}}, \kw{let}\ z = \kw{ctor}_i\ \many{y};\ C(F, \beta_l), \beta_l)$
C (let z = reuse x in ctor_i $\many{y}$; F, $\beta_l$) =
   $C_{\textit{app}}(\many{y}, \many{\own{}}, \kw{let}\ z =\ \kw{reuse}\ x\ \kw{in}\ \kw{ctor}_i\ \many{y};\ C(F, \beta_l), \beta_l)$

$C_{\textit{app}} : \ty{Var}^* \times \{\own{}, \bor{}\}^* \times \ty{FnBody}_{\textit{RC}} \times (\ty{Var} \rightarrow \{\own{}, \bor{}\}) \rightarrow \ty{FnBody}_{\textit{RC}}$
$C_{\textit{app}}(y\ \many{y'}, \own{}\ \many{b}, \kw{let}\ z = e;\ F, \beta_l) =$
      $\own{}^+_y(\many{y'} \cup \FV(F), C_{\textit{app}}(\many{y'}, \many{b}, \kw{let}\ z = e;\ F, \beta_l), \beta_l)$
$C_{\textit{app}}(y\ \many{y'}, \bor{}\ \many{b}, \kw{let}\ z = e;\ F, \beta_l) =$
      $C_{\textit{app}}(\many{y'}, \many{b}, \kw{let}\ z = e;\ \own{}^-_y(F, \beta_l), \beta_l)$
$C_{\textit{app}}([], \_, \kw{let}\ z = e;\ F, \beta_l) = \kw{let}\ z = e;\ F$
\end{lstlisting}
\caption{Inserting \kw{inc}/\kw{dec} instructions \label{fig:compiler}}
\end{figure}
\subsection*{Examples}
We demonstrate the behavior of the compiler on two application special cases.
The value of $\beta_l$ is constant in these examples and left implicit in applications.
\begin{enumerate}
\item Consuming the same argument multiple times
  \begin{align*}
    \beta(\var{c}) &:= \own{}\ \own{}\ \\
    \beta_l &:= [\var{y} \mapsto \own{}] \\
    C(\kw{let}\ &\var{z} = \var{c}\ \var{y}\ \var{y};\ \kw{ret}\ \var{z}) \\
             &= C_{\textit{app}}(\var{y}\ \var{y}, \own{}\ \own{}, \kw{let}\ \var{z} = \var{c}\ \var{y}\ \var{y};\ C(\kw{ret}\ \var{z})) \\
             &= C_{\textit{app}}(\var{y}\ \var{y}, \own{}\ \own{}, \kw{let}\ \var{z} = \var{c}\ \var{y}\ \var{y};\ \kw{ret}\ \var{z}) \\
             &= \own{}^+_\var{y}(\{\var{y}, \var{z}\}, C_{\textit{app}}(\var{y}, \own{}, \kw{let}\ \var{z} = \var{c}\ \var{y}\ \var{y};\ \kw{ret}\ \var{z})) \\
             &= \own{}^+_\var{y}(\{\var{y}, \var{z}\}, \own{}^+_\var{y}(\{\var{z}\}, C_{\textit{app}}([], [], \kw{let}\ \var{z} = \var{c}\ \var{y}\ \var{y};\ \kw{ret}\ \var{z}))) \\
             &= \own{}^+_\var{y}(\{\var{y}, \var{z}\}, \own{}^+_\var{y}(\{\var{z}\}, \kw{let}\ \var{z} = \var{c}\ \var{y}\ \var{y};\ \kw{ret}\ \var{z})) \\
             &= \own{}^+_\var{y}(\{\var{y}, \var{z}\}, \kw{let}\ \var{z} = \var{c}\ \var{y}\ \var{y};\ \kw{ret}\ \var{z}) \\
             &= \kw{inc}\ \var{y};\ \kw{let}\ \var{z} = \var{c}\ \var{y}\ \var{y};\ \kw{ret}\ \var{z}
  \end{align*}
  Because \var{y} is dead after the call, it needs to be incremented only once,
  \emph{moving} its last token to \var{c} instead.
  \item Borrowing and consuming the same argument
  \begin{align*}
    \beta(\var{c}) &:= \bor{}\ \own{}\ \\
    \beta_l &:= [\var{y} \mapsto \own{}] \\
    C(\kw{let}\ &\var{z} = \var{c}\ \var{y}\ \var{y};\ \kw{ret}\ \var{z}) \\
             &= C_{\textit{app}}(\var{y}\ \var{y}, \bor{}\ \own{}, \kw{let}\ \var{z} = \var{c}\ \var{y}\ \var{y};\ C(\kw{ret}\ \var{z})) \\
             &= C_{\textit{app}}(\var{y}\ \var{y}, \bor{}\ \own{}, \kw{let}\ \var{z} = \var{c}\ \var{y}\ \var{y};\ \kw{ret}\ \var{z}) \\
             &= C_{\textit{app}}(\var{y}, \own{}, \kw{let}\ \var{z} = \var{c}\ \var{y}\ \var{y};\ \own{}^-_\var{y}(\kw{ret}\ \var{z})) \\
             &= C_{\textit{app}}(\var{y}, \own{}, \kw{let}\ \var{z} = \var{c}\ \var{y}\ \var{y};\ \kw{dec}\ \var{y};\ \kw{ret}\ \var{z}) \\
             &= \own{}^+_\var{y}(\{\var{y}, \var{z}\}, \ C_{\textit{app}}([], [], \kw{let}\ \var{z} = \var{c}\ \var{y}\ \var{y};\ \kw{dec}\ \var{y};\ \kw{ret}\ \var{z})) \\
             &= \own{}^+_\var{y}(\{\var{y}, \var{z}\}, \kw{let}\ \var{z} = \var{c}\ \var{y}\ \var{y};\ \kw{dec}\ \var{y};\ \kw{ret}\ \var{z}) \\
             &= \kw{inc}\ \var{y};\ \kw{let}\ \var{z} = \var{c}\ \var{y}\ \var{y};\ \kw{dec}\ \var{y};\ \kw{ret}\ \var{z}
  \end{align*}
  Even though the owned parameter comes after the borrowed parameter, the presence of $y$ in the \kw{dec} instruction emitted when handling the first parameter makes sure we do not accidentally move ownership when handling the second parameter, but copy $y$ by emitting an \kw{inc} instruction.
\end{enumerate}

\subsection*{Preserving tail calls}
A \emph{tail call} $\kw{let}\ r = c\ \many{x};\ \kw{ret}\ r$ is an
application followed by a $\kw{ret}$ instruction. Recursive tail calls
are implemented using \emph{goto}s in our compiler backend. Thus, it
is highly desirable to preserve them as we transform \pureIR{} into
\rcIR{}. However, the previous example shows that our function for inserting reference
counting instructions may insert $\kw{dec}$ instructions after a constant application, and
consequently, destroy tail calls. A $\kw{dec}$ instruction is inserted after a constant application
$\kw{let}\ r = c\ \many{x}$ if $\beta(c)_i = \bor{}$ and $\beta_l(x_i) = \own{}$ for some
$x_i \in \many{x}$. That is, function $c$ takes the $i$-th
parameter as a borrowed reference, but the actual argument is owned. As an example,
consider the following function in \pureIR{}.
\begin{lstlisting}
  f x = case x of
    (let r = proj_1; ret r)
    (let y_1 = ctor_1; let y_2 = ctor_1 y_1; let r = f y_2; ret r)
\end{lstlisting}
The compiler from \pureIR{} to \rcIR{}, infers $\beta(f) = \bor{}$, and produces
\begin{lstlisting}
  f x = case x of
     (let r = proj_1 x; inc r; ret r)
     (let y_1 = ctor_1; let y_2 ctor_1 y_1;
      let r = f y_2; dec y_2; ret r)
\end{lstlisting}
which does not preserve the tail call $\kw{let}\ r = f\ y_2;\ \kw{ret}\ r$.
We addressed this issue in our real implementation by refining our
borrowing inference heuristic, marking $\beta(c)_i = \own{}$ whenever
$c$ occurs in a tail call $\kw{let}\ r = c\ \many{x};\ \kw{ret}\ r$
where $\beta_l(x_i) = \own{}$. This small modification guarantees that tail calls are
preserved by our compiler, and the following \rcIR{} code is produced for $f$ instead
\begin{lstlisting}
   f x = case x of
     (let r = proj_1 x; inc r; dec x; ret r)
     (dec x; let y_1 = ctor_1; let y_2 = ctor_1 y_1;
      let r = f y_2; ret r)
\end{lstlisting}


\section{Optimizing functional data structures for $\kw{reset}$/$\kw{reuse}$}
\label{sec:reuse}

In the previous section, we have shown how to automatically insert
$\kw{reset}$ and $\kw{reuse}$ instructions that minimize the number of
memory allocations at execution time. We now discuss techniques we have been using
for taking advantage of this transformation when writing functional
code. Two fundamental questions when using this optimization are: Does
a $\kw{reuse}$ instruction now guard my constructor applications?
Given a $\kw{let}\ y = \kw{reset}\ x$ instruction, how often is $x$
not shared at runtime? We address the first question using a simple
static analyzer that when invoked by a developer, checks whether
\kw{reuse} instructions are guarding constructor applications in a
particular function.  This kind of analyzer is straightforward to
implement in Lean since our IR is a Lean inductive datatype. This kind
of analyzer is in the same spirit of the \emph{inspection-testing}
package available for GHC~\cite{breitner2018promise}.
We cope with the second question using
runtime instrumentation.  For each $\kw{let}\ y = \kw{reset}\ x$
instruction, we can optionally emit two counters that track how often
$x$ is shared or not. We have found these two simple techniques quite
useful when optimizing our own code. Here, we report one instance
that produced a significant performance improvement.

\subsection*{Red-black trees}
Red-black trees are implemented in the Lean standard
library and are often used to write proof automation.
For the purposes of this section, it is sufficient to have
an abstract description of this kind of tree, and one of the re-balancing functions used by the insertion
function.
\begin{lstlisting}
Color  = R | B
Tree a = E | T Color (Tree a) a (Tree a)
balance_1 v t (T _ (T R l x r_1) y r_2)  = T R (T B l x r_1) y (T B r_2 v t)
balance_1 v t (T _ l_1 y  (T R l_2 x r)) = T R (T B l_1 y l_2) x (T B r v t)
balance_1 v t (T _ l y r)              = T B (T R l y r) v t
insert (T B a y b) x = balance_1 y b (insert a x) if x < y and a is red
...
\end{lstlisting}
Note that the first two $\textit{balance}_1$ equations create three $T$ constructor values, but the patterns on the
left-hand side use only two $T$ constructors. Thus, the generated IR for $\textit{balance}_1$ contains $T$ constructor
applications that are not guarded by $\textit{reuse}$, and this fact can be detected at compilation time.
Note that even if the result of $(\textit{insert}\ a\ x)$ contains only nonshared values, we still have to allocate one
constructor value. We can avoid this unnecessary memory allocation by inlining $\textit{balance}_1$.
After inlining, the input value $(T\ B\ a\ y\ b)$ is reused in the $\textit{balance}_1$ code.
The final generated code now contains a single constructor application that is not guarded by a $\textit{reuse}$, the one for
the equation:
\begin{lstlisting}
insert E x = T R E x E
\end{lstlisting}
The generated code now has the property that if the input tree is not shared, then only a single new node is allocated.
Moreover, even if the input tree is shared we have observed a positive performance impact using \kw{reset} and \kw{reuse}.
The recursive call $(\textit{insert}\ a\ x)$ always returns a nonshared node even if $x$ is shared. Thus,
$\textit{balance}_1\ y\ b\ (\textit{insert}\ a\ x)$ always reuses at least one memory cell at runtime.

There is another way to avoid the unnecessary memory allocation that does not rely on inlining.
We can chain the $T$ constructor value from $\textit{insert}$ to $\textit{balance}_1$.
We accomplish this by rewriting $\textit{balance}_1$ and $\textit{insert}$ as follows
\begin{lstlisting}
balance_1 (T _ _ v t) (T _ (T R l x r_1) y r_2)  = T R (T B l x r_1) y (T B r_2 v t)
balance_1 (T _ _ v t) (T _ l_1 y  (T R l_2 x r)) = T R (T B l_1 y l_2) x (T B r v t)
balance_1 (T _ _ v t) (T _ l y r)               = T B (T R l y r) v t
insert (T B a y b) x = balance_1 (T B E y b) (insert a x) if x < y and a is red
\end{lstlisting}
Now, the input value $(T\ B\ a\ y\ b)$ is reused to create value $(T\ B\ E\ y\ b)$ which is passed to $\textit{balance}_1$. Note that
we have replaced $a$ with $E$ to make sure the recursive application $(\textit{insert}\ a\ x)$
may also perform destructive updates if $a$ is not shared.
This simple modification guarantees that $\textit{balance}_1$ does not allocate memory when
the input trees are not shared.

\section{Runtime implementation}
\label{sec:runtime}


\subsection{Values}

In our runtime, every value starts with a header containing two
tags. The first tag specifies the value kind: \kw{ctor}, \kw{pap},
\kw{array}, \kw{string}, \kw{num}, \kw{thunk}, or \kw{task}. The
second tag specifies whether the value is single-threaded,
multi-threaded, or \emph{persistent}. We will describe how this kind is used to
implement thread safe reference counting in the next subsection. The kinds \kw{ctor} and
\kw{pap} are used to implement the corresponding values used in the
formal semantics of \pureIR{} and \rcIR{}.
The kinds \kw{array}, \kw{string}, and \kw{thunk} are self
explanatory.  The kind \kw{num} is for arbitrary precision numbers
implemented using the GNU multiple precision library (GMP).
The \kw{task} value is described in the next subsection.

Values tagged as single- or multi-threaded also contain a reference
counter. This counter is stored in front of the standard value header.
We will primarily focus on the layout of \kw{ctor} values here because  it is the most
relevant one for the ideas presented in this paper. A $\kw{ctor}_i$
value header also includes the constructor index $i$, the number of
pointers to other values and/or boxed values, and the number of bytes used to store scalar
unboxed values such as machine integers and enumeration types. In a
64-bit machine, the \kw{ctor} value header is 16 bytes long, twice the
size of the header used in OCaml to implement the corresponding kind
of value.  After the header, we store all pointers to other values
and boxed values, and then all unboxed values. Thus, in a 64 bit machine,
our runtime uses 32 bytes to implement a \emph{List} \emph{Cons} value:
16 bytes for the header, and 16 bytes for storing the list head and tail.
The unboxed value support has restrictions similar to the ones found in GHC.
For example, to pass an unboxed value to a polymorphic function we must first box it.

Our runtime has built-in support for array and string operations.
Strings are just a special case of arrays where the elements are
characters. We perform destructive updates when the array is not shared.
For example, given the array write primitive
\begin{lstlisting}
Array.write : Array $\alpha$ -> Nat -> $\alpha$ -> Array $\alpha$
\end{lstlisting}
the function application $\textit{Array.write}\ a\ i\ v$ will
destructively update and return the array $a$ if it is not shared.
This is a well known optimization for systems based on
reference counting~\cite{GCbook}, nonetheless we mention it here
because it is relevant for many applications. Moreover,
destructive array updates and our \kw{reset}/\kw{reuse} technique
complement each other. As an example, if we have a nonshared list of
integer arrays \textit{xs},
$\textit{map}\ (\textit{Array.map}\ \textit{inc})\ \textit{xs}$
destructively updates the list and all arrays. In the experimental
section we demonstrate that our pure quick sort is as efficient as the
quick sort using destructive updates in OCaml, and the quick sort
using the primitive ST monad in Haskell.




\subsection{Thread safety}

We use the following basic task management primitives to
develop the Lean frontend.
\begin{lstlisting}
Task.mk :$\ $ (Unit -> $\alpha$) -> Task $\alpha$
Task.bind : Task $\alpha$ -> ($\alpha$ -> Task $\beta$) -> Task $\beta$
Task.get : Task $\alpha$ -> $\alpha$
\end{lstlisting}
The function $\textit{Task.mk}$ converts a closure into a \kw{task} value
and executes it in a separate thread,
$\textit{Task.bind}\ t\ f$ creates a \kw{task} value that waits for $t$ to finish and
produce result $a$, and then starts $f\ a$ and waits for it to finish.
Finally, $\textit{Task.get}\ t$ waits for $t$ to finish and returns the
value produced by it. These primitives are part of the Lean runtime,
implemented in C++, and are available to regular users.

The standard way of implementing thread safe reference counting uses
memory fences~\cite{Boost}.  The reference counters are incremented
using an atomic fetch and add operation with a relaxed memory
order. The relaxed memory order can be used because new references to
a value can only be formed from an existing reference, and passing
an existing reference from one thread to another must already provide
any required synchronization. When decrementing a reference counter, it
is important to enforce that any decrements of the counter from other threads
are visible before checking if the object should be deleted. The
standard way of achieving this effect uses a \emph{release} operation
after dropping a reference, and an \emph{acquire} operation before
the deletion check. This approach has been used in the previous
version of the Lean compiler, and we have observed that the memory
fences have a significant performance impact even when only one thread
is being executed. This is quite unfortunate because most values
are only touched by a single execution thread.

We have addressed this performance problem in our runtime by tagging
values as \emph{single-threaded}, \emph{multi-threaded}, or
\emph{persistent}. As the name suggests, a single-threaded value is
accessed by a single thread and a multi-threaded one by one or more
threads. If a value is tagged as single-threaded, we do not use any
memory fence for incrementing or decrementing its reference counter.
Persistent values are never deallocated and do not even need a
reference counter.  We use persistent values to implement values
that are created at program initialization time and remain alive
until program termination.  Our runtime enforces the following
invariant: from persistent values, we can only reach other persistent
values, and from multi-threaded values, we can only reach persistent
or multi-threaded values. There are no constraints on the kind of
value that can be reached from a single-threaded value. By default,
values are single-threaded, and our runtime provides a
$\textit{markMT}(o)$ procedure that tags all single-threaded values
reachable from $o$ as multi-threaded.  This procedure is used to
implement $\textit{Task.mk}\ f$ and $\textit{Task.bind x f}$.  We use
$\textit{markMT}(f)$ and $\textit{markMT}(x)$ to ensure that all
values reachable from these values are tagged as multi-threaded \emph{before} we
create a new task, that is, while they are still accessible from only one thread.
Our invariant ensures that $\textit{markMT}$ does
not need to visit values reachable from a value already tagged
as multi-threaded.  Thus values are visited at most once by
$\textit{markMT}$ during program execution. Note that task creation is not
a constant time operation in our approach because it is proportional to the number of
single-threaded values reachable from $x$ and $f$. This does not seem
to be a problem in practice, but if it becomes an issue we can provide a
primitive $\textit{asMT}\ g$ that ensures that all values allocated
when executing $g$ are immediately tagged as multi-threaded.
Users would then use this flag in code that creates the values reachable by
$\textit{Task.mk}\ f$ and $\textit{Task.bind x f}$.

The reference counting operations perform an extra operation to test
the value tag and decide whether a memory fence is needed or not.
This additional test does not require any synchronization because the
tag is only modified before a value is shared with other execution
threads. In the experimental section, we demonstrate that this simple
approach significantly boosts performance. This is not surprising
because the additional test is much cheaper than memory fences on
modern hardware.
The approach above can be adapted to more complex libraries for
writing multi-threaded code. We just need to identify which functions
may send values to other execution threads, and use $\textit{markMT}$.


\section{Experimental evaluation}
\label{sec:evaluation}

We have implemented the RC optimizations described in the previous
sections in the new compiler for the Lean programming language.
We have implemented all optimizations in Lean, and they are available
online~\footnote{\url{https://github.com/leanprover/lean4/tree/master/library/init/lean/compiler/ir}}.
At the time of writing, the compiler supports only one backend where we emit
C code. We are currently working on an LLVM backend for
our compiler. To test the efficiency of the compiler and RC
optimizations, we have devised a number of benchmarks\footnote{\url{https://github.com/leanprover/lean4/tree/IFL19/tests/bench}}
that aim to replicate common tasks performed in compilers and proof assistants.
All timings are arithmetic means of 50 runs as reported by the \texttt{temci}
benchmarking tool~\cite{bechberger16bachelorarbeit},
executed on a PC with an i7-3770 Intel CPU and 16~GB RAM running Ubuntu 18.04,
using Clang 9.0.0 for compiling the Lean runtime library
as well as the C code emitted by the Lean compiler.
\begin{itemize}
\item \verb!deriv! and \verb!const_fold! implement
  differentiation and constant folding, respectively, as examples of symbolic
  term manipulation where big expressions are constructed and transformed.
  We claim they reflect operations frequently performed by proof automation procedures used in theorem provers.
\item \verb!rbmap! stress tests the red-black tree implementation from the Lean standard library.
  The benchmarks \verb!rbmap_10! and \verb!rbmap_1! are two variants where
  we perform updates on shared trees.
\item \verb!parser! is the new parser we are developing for the next version of
  Lean. It is written purely in Lean (approximately $2000$ lines of code).
\item \verb!qsort! it is the basic quicksort algorithm for sorting arrays.
\item \verb!binarytrees! is taken from the Computer Languages Benchmarks
Game\footnote{\url{https://benchmarksgame-team.pages.debian.net/benchmarksgame/performance/binarytrees.html}}.
This benchmark is a simple adaption of Hans Boehm's \texttt{GCBench} benchmark\footnote{\url{http://hboehm.info/gc/gc_bench/}}.
The Lean version is a translation of the fastest, parallelized Haskell solution,
using \texttt{Task} in place of the Haskell \texttt{parallel} API.
\item \verb!unionfind! implements the union-find algorithm which is frequently used
  to implement decision procedures in automated reasoning. We use arrays to store
  the \emph{find} table, and thread the state using a state monad transformer
\end{itemize}

We have tested the impact of each optimization by selectively disabling it and
comparing the resulting runtime with the base runtime~(\cref{fig:leanbench}):
\begin{itemize}
\item \emph{-reuse} disables the insertion of \kw{reset}/\kw{reuse} operations
\item \emph{-borrow} disables borrow inference, assuming that all parameters are
  owned. Note that the compiler must still honor borrow annotations on builtins,
  which are unaffected.
\item \emph{-ST} uses atomic RC operations for all values
\end{itemize}

\begin{figure}
  \centering
  \begin{tabular}{lrrrr} \toprule
    & base & \emph{-reuse} & \emph{-borrow} & \emph{-ST} \\ \midrule
    \input{report_lean}
    \bottomrule
  \end{tabular}
  \caption{Lean variant benchmarks, normalized by the base run time (\texttt{rbmap} for \texttt{rbmap\_*}). Digits whose
    order of magnitude is no larger than that of twice the standard deviation are marked by squiggly lines.}
  \label{fig:leanbench}
\end{figure}

The results show that the new \kw{reset} and \kw{reuse} instructions significantly improve
performance in the benchmarks \verb!const_fold!, \verb!rbmap!, and
\verb!unionfind!. The borrowed inference heuristic
provides significant speedups in benchmarks \verb!binarytrees! and
\verb!deriv!.

We have also directly translated some of these programs to other statically
typed, functional languages: Haskell, OCaml, and Standard ML (\cref{fig:crossbench}).
For the latter we selected the compilers MLton~\cite{Weeks:2006:WCM:1159876.1159877},
which performs whole program optimization and can switch between multiple GC schemes at runtime,
and MLKit, which combines Region Inference and garbage collection~\cite{het02}.
While not primarily a functional language, we have also included Swift as a
popular statically typed language using reference counting.
For \verb!binarytrees!, we have used the original files and compiler flags from the fastest Benchmark
Game implementations. For Swift, we used the second-fastest, safe
implementation, which is much more comparable to the other versions than the
fastest one completely depending on unsafe code. The Benchmark Game does not include an SML version.
For \verb!qsort!, the Lean code is pure and relies on the fact that array
updates are destructive if the array is not shared. The Swift code behaves
similarly because Swift arrays are copy-on-write.
All other versions use destructive updates, using the \textit{ST} monad in the case of Haskell.

\begin{figure*}
  \centering
  \begin{tabular}{lrrrrrrrrrrrrrrrrrr} \toprule
    &\multicolumn{3}{c}{Lean 4} & \multicolumn{3}{c}{GHC 8.8.3} & \multicolumn{3}{c}{ocamlopt 4.10} & \multicolumn{3}{c}{MLton 20180207} & \multicolumn{3}{c}{MLKit 4.4.2} & \multicolumn{3}{c}{Swift 5.1.1} \\
    \cmidrule(lr{.5em}){2-4} \cmidrule(lr{.5em}){5-7} \cmidrule(lr{.5em}){8-10} \cmidrule(lr{.5em}){11-13} \cmidrule(lr{.5em}){14-16} \cmidrule(lr{.5em}){17-19}
    & Time & Del & CM & Time & GC & CM & Time & GC & CM & Time & GC & CM & Time & GC & CM & Time & GC & CM \\ \midrule
    \input{report_cross}
    \bottomrule
  \end{tabular}
  \footnotetext{Because OCaml does not currently support multi-threading, the
    program uses the Unix \texttt{fork} call instead}

  \caption{Cross-language benchmarks. The measurements include wall clock time
    (normalized by the Lean base run time), GC time (in percent, as reported
    by the respective compiler), and last-level cache misses (CM, in million per
    second, as reported by \texttt{perf stat}). For Swift, we measure time spent
    in inc, dec, and deallocation runtime functions as GC time using \texttt{perf}.
    For Lean, the former are always inlined, so we can only measure object deletion
    time.}
  \label{fig:crossbench}
\end{figure*}
While the absolute runtimes in \cref{fig:crossbench}
are influenced by many factors other than the implementation
of garbage collection that make direct comparisons difficult,
the results still signify that both our garbage collection
and the overall runtime and compiler implementation are very
competitive.  We initially conjectured the good performance was a result of
reduced cache misses due to reusing allocations and a lack
of GC tracing.
However, the results demonstrate this is not the case.  The
only benchmark where the code generated by our compiler produces
significantly fewer cache misses is \verb!rbmap!.  Note that Lean is
5x as fast as OCaml on \verb!const_fold!  even though it
triggers more cache misses per second.  The results
suggest that Lean code is often faster in the benchmarks where
the code generated by other compilers spends a significant amount of time performing GC. Using
\verb!const_fold! as an example again, Lean spends only $17\%$
of the runtime deallocating memory, while OCaml spends $90\%$ in the
GC.  This comparison is not entirely precise since it does not include
the amount of time Lean spends updating reference counts, but it seems
to be the most plausible explanation for the difference in
performance.  The results for \verb!qsort! are surprising, the
Lean and Swift implementations outperforms all destructive ones but
MLton. We remark that MLton and Swift have a clear advantage since they use
arrays of unboxed machine integers, while Lean and the other compilers
use boxed values. We did not find a way to disable this optimization
in MLton or Swift to confirm our conjecture.  We believe this benchmark
demonstrates that our compiler allows programmers to write efficient
pure code that uses arrays and hashtables. For \verb!rbmap!, Lean is
much faster than all other systems except for OCaml. We imagined this would only be the case
when the tree was not shared. Then we devised the two variants
\verb!rbmap_10! and \verb!rbmap_1! which save the
current tree in a list after every tenth or every insertion, respectively.
The idea is to simulate the behavior of a backtracking search where we
store a copy of the state before each case-split. As expected, Lean's
performance decreases on these two variants since the tree is now a
shared value, and the time spent deallocating objects increases
substantially.  However, Lean still outperforms all systems but MLton
on \verb!rbmap_1!.  In all other systems but MLton and Swift, the time
spent on GC increases considerably. Finally, we point out that
MLton spends significantly less time on GC than the other languages using
a tracing GC in general.



\section{Related work}
\label{sec:related}

The idea of representing RC operations as explicit instructions so as to
optimize them via static analysis is described as early as
\citet{Barth:1977:SGC:359636.359713}.
\citet{Schulte:1994} describes a system with many
features similar to ours. 
In general, Schulte's language is much simpler than ours, with a single list type
as the only non-primitive type, and no higher-order functions. He does
not give a formal dynamic semantics 
for his system. He gives an algorithm for inserting RC instructions that, like ours,
has an on-the-fly optimization for omitting \kw{inc} instructions if a
variable is already dead and would immediately be decremented
afterwards. Schulte briefly discusses how RC operations can be
minimized by treating some parameters as ``nondestructive'' in the
sense of our borrowed references. In contrast to our inference of
borrow annotations, Schulte proposes to create one copy of a function
for each possible destructive/nondestructive combination of parameters
(i.e.\ exponential in the number of (non-primitive) parameters) and
to select an appropriate version for each call site of the function. Our
approach never duplicates code.

Introducing destructive updates into pure programs has traditionally focused on
primitive operations like array updates~\cite{Hudak:1985:AUP:318593.318660}, particularly in the functional array
languages \textsc{Sisal}~\cite{sisal} and \textsc{SaC}~\cite{sac}. \citet{sac-reuse} propose an \verb!alloc_or_reuse!
instruction for \textsc{SaC} that can select one of multiple array candidates for reuse, but do not describe
heuristics for when to use the instruction.
\citet{Ferey:2016:CGU:2963372.2963387} describe how functional update operations
explicit in the source language can be turned into destructive updates using the
reference counter. In contrast, \citet{Schulte:1994} presents a ``reusage'' optimization that has an effect similar to the one
obtained with our \kw{reset}/\kw{reuse} instructions. In particular, it is
independent of a specific surface-level update syntax.
However, his optimization (transformation $T14$) is more restrictive
and is only applicable to a branch of a $\kw{case}\ x$ if $x$ is dead at the beginning of the branch.
His optimization cannot handle the simple $\textit{swap}$ described earlier,
let alone more complex functions such as the red black
tree re-balancing function $\textit{balance}_1$.

While not a purely functional language, the Swift programming
language\footnote{\url{https://developer.apple.com/swift/}} has directly
influenced many parts of our work. To the best of our knowledge, Swift was the
first non-research language to use an intermediate representation with explicit RC
instructions, as well as the idea of (safely) avoiding RC operations via ``borrowed''
parameters (which are called ``+0'' or ``guaranteed'' in Swift), in its
implementation. While Swift's primitives may also elide copies when given a
unique reference, no speculative destructive updates are introduced for
user-defined types, but this may not be as important for an impure language as
it is for Lean. Parameters default to borrowed in Swift, but the compiler may
locally change the calling convention inside individual modules.

\citet{Baker:1994:MRC:185009.185016} describes optimizing reference counting by
use of \emph{two} pointer kinds, a standard one and a \emph{deferred increment}
pointer kind. The latter kind can be copied freely without adding RC operations,
but must be converted into the standard kind by incrementing it before storing it in an object or returning it. The two kinds
are distinguished at runtime by pointer tagging. Our borrowed references
can be viewed as a static refinement of this idea. Baker then describes an
extended version of deferred-increment he calls \emph{anchored} pointers that
store the stack level (i.e.\ the lifetime) of the standard pointer they have been
created from. Anchored pointers do not have to be converted to the standard kind if returned
from a stack frame above this level. In order to statically approximate this
extended system, we would need to extend our type system
with support for some kind of \emph{lifetime annotations} on return types
as featured in Cyclone~\cite{jim2002cyclone} and
Rust~\cite{Matsakis:2014:RL:2663171.2663188}.



\citet{dynamic-atomicity} optimize Swift's reference counting scheme
by using a single bit to tag objects possibly shared between multiple
threads, much like our approach. However, because of mutability, every
single store operation must be intercepted to (recursively) tag objects
before becoming reachable from an already tagged object.
\citet{biasedRC} remove the need for tagging by extending every object header
with the ID of the thread $T$ that allocated the value, and two
reference counters: a shared one that requires atomic operations, and
another one that is only updated by $T$.
Thanks to immutability, we can make use of the simpler scheme without
introducing store barriers during normal code generation. Object tagging
instead only has to be done in threading primitives.


\section{Conclusion}

We have explored reference counting as a memory management technique
in the context of an eager and pure functional programming
language. Our preliminary experimental results are encouraging and
show our approach is competitive with state-of-the-art compilers for
functional languages and often outperform them. Our resurrection
hypothesis suggests there are many opportunities for reusing memory
and performing destructive updates in functional programs.
We have also explored optimizations for reducing the number of reference counting updates,
and proposed a simple and efficient technique for implementing thread safe reference counting.

We barely scratched the surface of the design space, and there are many
possible optimizations and extensions to explore. We hope our \pureIR{} will be useful in the future
as a target representation for other purely functional languages (e.g.,
Coq, Idris, Agda, and Matita).
We believe our approach can be extended to programming
languages that support cyclic data structures because it is orthogonal
to traditional cycle-handling techniques. Finally, we are working on a
formal correctness proof of the compiler described in this paper,
using a type system based on intuitionistic linear logic to model owned
and borrowed references.

\begin{acks}
  We are very grateful to
  Thomas Ball,
  Johannes Bechberger,
  Christiano Braga,
  Sebastian Graf,
  Simon Peyton Jones,
  Daan Leijen,
  Tahina Ramananandro,
  Nikhil Swamy,
  Max Wagner
  and the anonymous reviewers
  for extensive comments, corrections and advice.
\end{acks}

\bibliography{refcount}
\end{document}